\documentclass[%
 reprint,
superscriptaddress,
 amsmath,amssymb,
 aps,
]{revtex4-2}

\usepackage{graphicx}
\usepackage{dcolumn}
\usepackage{bm}

\begin{document}

\preprint{APS/123-QED}

\title{Non-linear optical response of the excited Laughlin liquid}

\author{L. V. Kulik}
 \email{To whom correspondence should be addressed: kulik@issp.ac.ru}
\affiliation{Institute of Solid State Physics Russian Academy of Sciences Chernogolovka, Moscow District, 2 Academician Ossipyan Street, 142432, Russia}

\author{A. S. Zhuravlev} 
\affiliation{Institute of Solid State Physics Russian Academy of Sciences Chernogolovka, Moscow District, 2 Academician Ossipyan Street, 142432, Russia}

\author{A. V. Larionov}
\affiliation{Institute of Solid State Physics Russian Academy of Sciences Chernogolovka, Moscow District, 2 Academician Ossipyan Street, 142432, Russia}

\author{A. B. Van’kov}
\affiliation{Institute of Solid State Physics Russian Academy of Sciences Chernogolovka, Moscow District, 2 Academician Ossipyan Street, 142432, Russia}
\affiliation{National Research University Higher School of Economics, Moscow, 20 Myasnitskaya Street, 101000, Russia}

\author{A. A. Zagitova}
\affiliation{Institute of Solid State Physics Russian Academy of Sciences Chernogolovka, Moscow District, 2 Academician Ossipyan Street, 142432, Russia}

\author{I. V. Kukushkin}
\affiliation{Institute of Solid State Physics Russian Academy of Sciences Chernogolovka, Moscow District, 2 Academician Ossipyan Street, 142432, Russia}

\author{V. Umansky}
\affiliation{Braun Center for Submicron Research, Weizmann Institute of Science, 234 Herzl Street, POB 26, Rehovot 76100, Israel }

\begin{abstract}
An ensemble of neutral excitations is constructed experimentally in the Laughlin liquid at the electron filling factor 1/3. The excitations are found to induce a nonlinear optical response, manifested as a quadratic dependence of the reflection signal on the excitation power. The reported experimental results indicate that the observed effect is due to the contribution of the coherent anti-Stokes-Stokes scattering of light from the excited Laughlin liquid.
\end{abstract}

\maketitle


A 2D electron system can be used to realize a variety of particle statistics. In three dimensions, the electron complexes can be either fermionic, including electrons themselves, holes, trions, plasmarons, etc, or bosonic, involving excitons, plasmons, biexcitons, plexcitons, etc. By contrast, a 2D electron system subjected to an external magnetic field permits a range of anyonic statistics of quasiparticles, falling in between the bosonic and fermionic limits \cite{wilczek1982}. The possibility of experimentally realizing anyonic statistics in two-dimensional electron systems was acknowledged soon after Laughlin’s pioneering work describing the electron system properties for several fractional quantum Hall-effect (FQHE) states \cite{laughlin1983anomalous}. However, only recently has it been proven that certain quasiparticles in two-dimensional electron systems are indeed anyons. In an FQHE state 1/3, the behavior of charged quasiparticles was found to be unlike that with bosonic or fermionic statistics. Instead, they behave as true anyons with statistics $\pi/3$ \cite{bartolomei2020fractional,nakamura2020direct}.  Moreover, more complicated statistics including the non-abelian kinds are expected in FQHE states \cite{willett2009measurement}.

The question arises as to what the neutral excitations are in the bulk of such anyonic matter. There is a theoretical consensus regarding magnetorotons – the neutral excitations with an orbital momentum of unity \cite{girvin1985collective} – that are assumed to be bosons. Yet, working with magnetorotons experimentally is a complex problem since they are well-defined only at large 2D momenta, on the order of the inverse magnetic length. At zero momentum (in the case most appropriate for an experimental study), these excitations fall into the multi-roton continuum and decay \cite{kukushkin2009dispersion,kang2000inelastic}. Nevertheless, the theoretical analysis predicts another low-energy boson  branch of neutral excitations capable of surviving at zero momentum – namely, the “magnetograviton” branch \cite{haldane2011geometrical,gromov2017geometrical,can2018geometrical}.

Magnetogravitons are some of the most exotic quasiparticles in solid-state physics. Conventionally, they are described in terms of the perturbations of the space metric introduced for a system of FQHE quasiparticles \cite{haldane2011geometrical}. They have an orbital momentum of two. If achieved experimentally, these excitations would rapidly decay \cite{wiegmann2018time}. However, it has been found that given certain parameters of the 2D system confining potential, the analogues of magnetogravitons with the unit spin (spin-magnetogravitons) can be the lowest-energy, zero-momentum neutral excitations. The authors of the paper have been able to create the required experimental conditions to collect an ensemble of spin-magnetogravitons \cite{kulik2021laughlin}. In the present work, we investigate the optical properties of the Laughlin liquid excited with spin-magnetogravitons and observe a new effect – a nonlinear optical response of the excited Laughlin liquid.

\begin{figure}[b]
\includegraphics{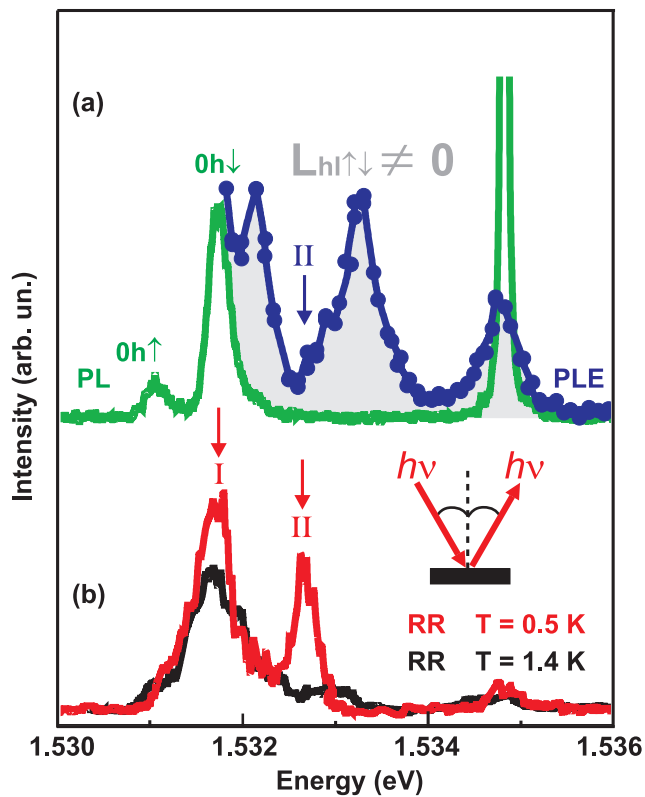}
\caption{\label{fig:fig1} (a) PLE (blue dots) and resonant PL (green line) of the Laughlin liquid, measured at the temperature of 0.5 K. Allowed optical transitions 0e-0h are marked as 0h$\uparrow$ and 0h$\downarrow$ depending on the electron spin in the conductance band of GaAs (the electron $ g$-factor in GaAs is negative).  The gray background marks the region of weakly allowed optical transitions from the valence band to the conduction band, from the hole states with predominantly non zero orbital angular momentum. (b) The resonant reflection (RR) spectrum of the Laughlin liquid at 0.5 K (red line), and of the electron system at 1.4 K, when the Laughlin liquid is destroyed (black line). The inset shows the scheme for measuring the RR signal.}
\end{figure}

\begin{figure}[b]
\includegraphics{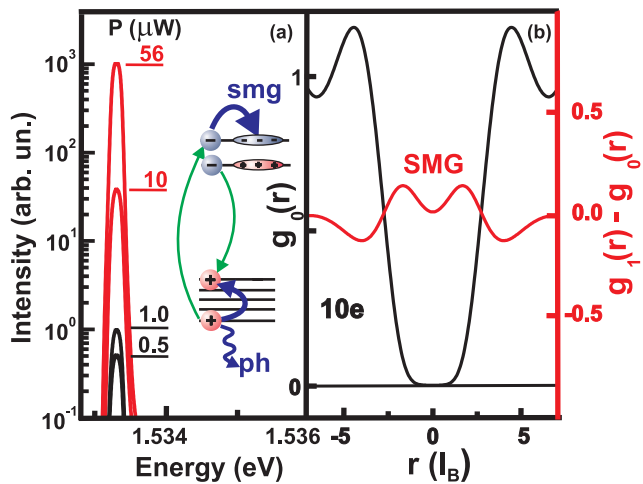}
\caption{\label{fig:fig2} (a) RR intensity obtained at the line II energy at the power levels from 0.5 to 56 $\mu$W. The line width is defined by the spectrometer normal slit. The diagram depicts the schemes of optical transitions that form spin-magnetogravitons (smg). (b) Left vertical axis – the pair correlation function $g(r)$ for the ground state of the Laughlin liquid at the electron filling factor 1/3 (black curve). Right vertical axis – the difference between the pair correlation functions for the excited and ground states plotted for the zero-momentum spin-magnetograviton (red solid line). The variable $r$ is given in the magnetic-length units. The numerical calculation is carried out for 10 electrons, following the procedure in \cite{kulik2021laughlin}.}
\end{figure}

In our experiments, an ensemble of spin-magnetogravitons was created using an 18-nm-wide quantum well with the electron concentration of $8.4\cdot 10^{10}$cm$^{-2}$. The sample was placed into a cryostat with pumped 3He vapors to provide the lowest sample temperature of 0.5 K. The optical measurements were taken at the external magnetic field of up to 14 T in two different experimental configurations. The two-fiber configuration (I) had one fiber delivering the pump laser radiation to the sample, while the other was used to collect the reflected light and guide it to the entrance slit of a spectrometer. The optical-window configuration (II) had a single glass window used both to conduct the excitation light to the sample and to collect the scattered light. The uniformly excited area in the first experimental configuration was 600 microns in diameter, whereas the second experimental configuration allowed to focus the pump laser beam into a much smaller spot (less than 100 microns in diameter), though with the non-uniform spatial distribution. A spectroscopic technique based on the photo-reflection of light (RR) was employed (Fig.~\ref{fig:fig2}). The electron system was excited using a tunable, continuous-wave (CW) laser source with a linewidth of 10 MHz. For measuring the resonant reflectance spectra, the experimental geometry was chosen so that the specularly reflected beam axis coincided with the receiving fiber axis at an angle of incidence of approximately 10$^\circ$. The contribution of the sample surface reflection was suppressed using crossed linear polarizers set in the exciting and reflecting laser beams. Because of the broken time-reversal symmetry for the electron system in the external magnetic field, all of its allowed optical transitions are active in circularly polarized light. Therefore, the linear polarized photons scattered from the electron system easily pass through the second crossed polarizer, whereas the unwanted surface scattering does not change the linear polarization of light and thus cannot pass through the second polarizer. We have also measured the resonant photoluminescence (PL) and photoexcitation (PLE) spectra of the electron system (Fig.~\ref{fig:fig1}). 

In the PLE spectrum of the examined quantum well, we observe the lines corresponding to the allowed two-particle optical transitions from the valence band to the zero Landau level of the conduction band. The hole states from the light- and heavy-hole zones of the valence band in the magnetic field are the superpositions of the states from the Landau levels 0, 1, 2, 3 and the spin projections onto the axis of the magnetic field -3/2, -1/2, 1/2, and 3/2. The merging of the finite-width Landau levels results in a continuous region of weakly allowed optical transitions, with energies above the transition energies of the allowed optical transitions from the zero Landau level of heavy holes to the zero Landau level of the conduction band (Fig.~\ref{fig:fig1}). Consequently, in the resonant photoluminescence (PL) spectra, the significant oscillation strength is manifested by the transitions between the zero Landau levels  (0e-0h). Since Laughlin liquid occupies the lowest spin sublevel of the zero electron Landau level, its resonant PL intensity is substantially less compared to the PL intensity from the upper spin sublevel \cite{kulik2021laughlin}. For the same reason, in the RR spectrum of the electron system, we observe only a single prominent line (I) with an intensity significantly exceeding those of the other RR lines. The physics of the reflection channel-I becomes clear from the corresponding PL spectrum. It can be outlined as follows: the absorption of a photon from the laser source creating an electron-hole pair with a hole from the zero Landau level of the heavy hole subband of the valence band and an electron from the empty upper spin sublevel of the zero Landau level of the conduction band followed by the  recombination of the photoexcited electron-hole pair.

Besides the major line (I) in the RR spectrum of the Laughlin liquid, we observe a line (II) of appreciable intensity associated with spin-magnetogravitons in the  Laughlin liquid, whose energy does not correspond to a maximum in the density of the valence band states (Fig.~\ref{fig:fig1}) \cite{kulik2021laughlin}. The intensity of the line (II) varies greatly when additional laser radiation with the energy exceeding that of the (0e-0h) optical transition forms an ensemble of ultra-long-life spin-magnetogravitons. 

Spin-magnetogravitons are formed through relaxation processes  (Fig.~\ref{fig:fig2}). An electron from the valence band of a quantum well is transferred into the unoccupied upper spin-sublevel of the zero Landau level by means of a weakly allowed optical transition. A photoexcited valence band hole relaxes into the upper spin-sublevel of the zero heavy-hole Landau level due to strong spin-orbit coupling in the GaAs valence band, causing the change in the total spin of the electron system. Then, in the optical recombination with a Laughlin-liquid electron, the valence band hole transforms into a Fermi hole, taking place of an electron on the lowest spin sublevel of the zero Landau level of the conduction band. As a result, a photoexcited electron and a Fermi hole can form a spin-magnetograviton provided that in the process of relaxation of the photoexcited hole in the valence band, part of the relaxation energy necessary for exciting the spin-magnetograviton is transferred to the Laughlin-liquid electrons (Fig.~\ref{fig:fig2}). In our study, we found that the largest growth rate of line II per unit of power of additional laser radiation occurs when the laser radiation has the same energy as the optical transition associated with the line II \cite{kulik2021laughlin}. Therefore, in this particular optical transition, the spin-magnetogravitons are formed most efficiently, whereas the most effective channel for creating spin-magnetogravitons is where all the relaxation energy in the valence band is transferred to the electron excitation in the Laughlin liquid. In that case, no extra acoustic phonon is needed for dissipating the excess energy (Fig.~\ref{fig:fig2}). This process is the resonant Stokes Raman scattering of light creating spin-magnetogravitons.

However, Stokes Raman scattering itself cannot lead to the appearance of a line in the reflection spectrum. To produce line II, the inverse anti-Stokes Raman scattering of light from the excited Laughlin liquid is required. It is not correct anymore to consider a “Fermi hole” in the Laughlin liquid that could be filled with a valence electron in the anti-Stokes Raman scattering of light, as such an excitation does not exist. As shown in Fig.~\ref{fig:fig2}, the positive and negative charges in a spin-magnetograviton are “spread out” in space over at least ten magnetic lengths. Nevertheless,   an optical transition from the valence band into the conduction band with the annihilation of the spin-magnetograviton is possible as a reverse process to the Stokes Raman scattering. Thus, two scattering processes may contribute to line II: i) the Stokes Raman scattering of light creating a spin-magnetograviton in the ground state of the Laughlin liquid, followed by the anti-Stokes Raman scattering annihilating that spin-magnetograviton (Stokes-anti-Stokes Raman scattering of light); ii) the anti-Stokes Raman scattering of light in the excited state of the Laughlin liquid annihilating an existing spin-magnetograviton, followed by the Stokes Raman scattering with the subsequent creation of an identical spin-magnetograviton (anti-Stokes-Stokes Raman scattering of light). Importantly, both processes must be coherent to contribute to the resonance reflection with zero energy (momentum) transfer. 

\begin{figure}[b]
\includegraphics{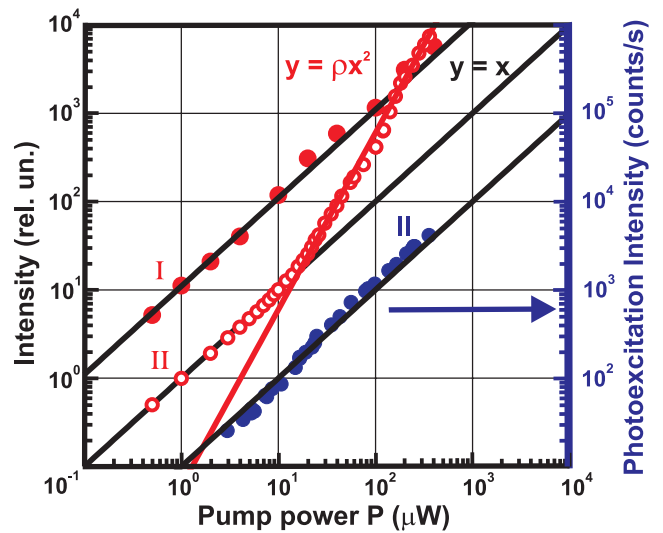}
\caption{\label{fig:fig3} Left axis – the dependence of the RR intensity on the laser excitation power obtained for the scattering channels I (red solid dots) and II (red empty circles). The intensity of the reflected signal is normalized by the proportionality coefficient $\alpha$ from Eq.~(\ref{eq:eq1}). Black lines are linear functions. The red line is the square function of the excitation power.  Right axis – the PLE intensity measured at the energy of line (II).  }
\end{figure}

\begin{figure}[b]
\includegraphics{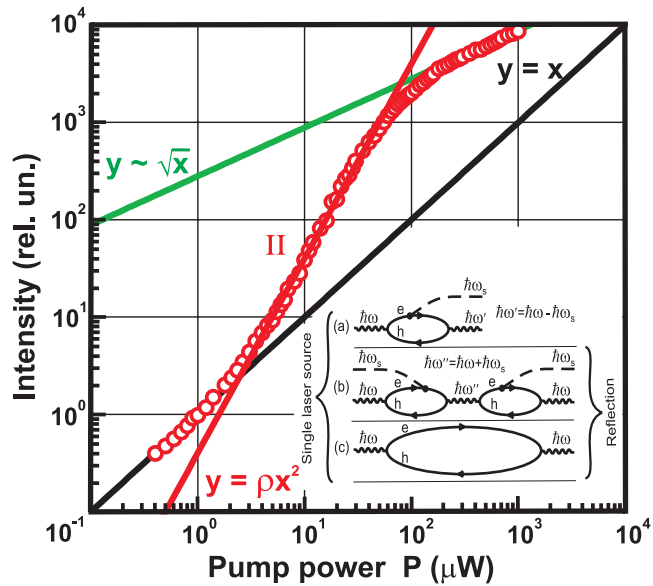}
\caption{\label{fig:fig4} The RR light intensity in channel-II versus the laser excitation power (empty circles), measured in the small laser excitation spot as described in the text. The intensity of the reflected signal is normalized by the proportionality coefficient $\alpha$ from Eq.~(\ref{eq:eq1}). The green, black, and red lines are square root, linear, and square functions of the excitation power, correspondingly. In the inset, the scheme of three optical processes stimulated by the laser photon with the energy of the line (II): (a) resonant Stokes Raman scattering of light, (b) coherent Anti-Stokes-Stokes Raman scattering of light, (c) resonant elastic backscattering of light (“idle” process).}
\end{figure}

Given the energy of line II, the reflection signal intensity is determined not only by the two aforementioned processes but also by the “idle” process – the resonance elastic backscattering from the Laughlin liquid. The coherent Stokes-anti-Stokes Raman scattering of light cannot be distinguished from the background “idle” process. However, the reverse anti-Stokes-Stokes Raman scattering of light is expected to cause a nonlinear dependence of the RR signal on the excitation radiation power if a single laser source is used for two processes: exciting spin-magnetogravitons and mesuring the light scattering from the existing spin-magnetogravitons. Indeed, when we use a single laser source with photon energy equal to that of the optical transition of line II, the scattering in channel-II starts to exhibit a superlinear (quadratic) enhancement on reaching a certain threshold of the laser excitation power (Fig.~\ref{fig:fig2}-\ref{fig:fig4}). 

The nonlinearity of the reflection signal intensity is determined by the two processes depicted in diagrams (a) and (b) in Fig.~\ref{fig:fig4}. Namely, the Stokes scattering (a) and the coherent anti-Stokes-Stokes scattering (b).  The optical processes considered in (b) and in the “idle” process (c) can be described by the following equation: 
\begin{eqnarray}
I = \alpha P + \beta Pn ~,
\label{eq:eq1}
\end{eqnarray}
where $I$ is the reflected light intensity, $P$ is the laser excitation power, and $n$ is the total number of spin-magnetogravitons in the Laughlin liquid. The first and the second terms in the sum are the respective contributions from (c) and (b). The $\alpha$ coefficient denotes the probability of light scattering in (c). It also accounts for the capability of our experimental setup to excite the electron system and collect the reflected light. Therefore, since the exact value of $\alpha$ is basically unknown, only the normalized quantity $\frac{I}{\alpha}$ has physical sense.

Carrying out standard calculations for the average number of excitations created in  (a) under the conditions of stationary excitation, we arrive at:
\begin{eqnarray}
\frac{dn}{dt} = \gamma P - \frac{n}{\tau} = 0 ~,
\label{eq:eq2}
\end{eqnarray}
where $\tau$ is the lifetime of spin-magnetogravitons and the coefficient $\gamma$ represents the probability of exciting a zero-momentum spin-magnetograviton in the process (a) taking into account all the unknowns of our experimental setup. Thus, we obtain:
\begin{eqnarray}
\frac{I}{\alpha} = P + \frac{\beta \gamma \tau}{\alpha}P^2 = P + \rho P^2 ~,
\label{eq:eq3}
\end{eqnarray}
with $\rho = \frac{\beta \gamma \tau}{\alpha}$, which yields the desired $P^2$ dependency (Fig.~\ref{fig:fig3}-\ref{fig:fig4}).

The resultant equations properly describe the dependence of the scattering signal intensity on the excitation power. At the initial stage, when the number of spin-magnetogravitons in the electron system is small, the “idle” process (c) dominates in the reflection spectrum. At the same time, the photons that excite the electron system are engaged in process (a), which leads to creating spin-magnetogravitons in the Laughlin liquid. Further increase in the excitation power activates the scattering channel (b), which is governed by the number of spin-magnetogravitons in the Laughlin liquid. As a result, we observe a quadratic dependency of the reflection signal (Fig.~\ref{fig:fig3}).

To exclude possible alternative processes that can lead to a nonlinear response in the reflection spectrum, we consider the well-established mechanisms of nonlinearity in multi-particle systems. There are two well-known effects that stimulate nonlinear light scattering, which are most prominent in atomic condensates \cite{ketterle2001collective}. The “material gain” is related to the bosonic statistics of excited states. Namely, when a scattering process with the creation of an excitation involves a quantum state already occupied with analogous excitations, the scattering is enhanced due to the Bose factor $N+1$, where $N$ is the number of excitations occupying this quantum state. In that case, the light scattering nonlinearity can be observed in light absorption also. Hence, the dependence of the scattered light intensity on the excitation power is expected to be more complicated than a quadratic relation. However, in our study, there is no evidence of deviation from the quadratic dependence over a wide range of excitation powers.

We independently measured the PLE intensity at the energy of line II as a function of the photoexcitation power. In high-mobility \mbox{AlGaAs/GaAs} quantum wells, the dependence of the PLE intensity on the excitation power is expected to be the same as the absorption intensity since the nonradiative decay time for the photoexcited electrons is much longer than the recombination time \cite{kulik1997effect}. The PLE intensity is found to be a linear function of the  photoexcitation power   (Fig.~\ref{fig:fig3}). Therefore, the observed nonlinear optical response from the excited Laughlin liquid is unrelated to the Bose statistics of excitations. At the same time, it cannot be attributed to the macro-filling of a particular photon mode involved in the scattering process because the given range of cw laser excitation power densities is too small for registering such an effect \cite{ketterle2001collective}.

As soon as the number of spin-magnetogravitons reaches the maximum allowed value (full saturation of the Laughlin liquid by excitations), we should see the saturation of process (a) and  (b); i.e. a further increase in the intensity of laser photoexcitation power should not lead to nonlinear growth in the intensity of the RR signal. Indeed, as the laser excitation spot is substantially reduced in size, and thus the excitation power density is increased, we observe that RR signal behavior changes from a quadratic to a sublinear dependency (Fig.~\ref{fig:fig4}).

In conclusion, we have discovered a quadratic component in the dependency of the reflection signal collected from the excited Laughlin liquid at the electron filling factor 1/3. The experimental data indigates that this contribution is due to the coherent anti-Stokes-Stokes scattering of light induced by a quasi-equilibrium ensemble of spin-magnetogravitons, the neutral excitations with spin one. A similar light scattering process should also be observed in other material systems. However, the ratio of the signals from the coherent anti-Stokes-Stokes scattering of light and from the elastic backscattering of light should be so small that it would be a very difficult experimental problem to distinguish the optical signal of the first process against the background of the second. The broken time-reversal symmetry for the Laughlin liquid, the ultra-long lifetimes of spin-magnetogravitons, and the resonance conditions associated with the mixing of the Landau levels of the heavy- and light-hole bands in GaAs/AlGaAs quantum wells create a unique opportunity for the direct observation of this interesting physical phenomenon. For the Laughlin liquid, this effect makes it possible to measure the exact zero-momentum spin-magnetograviton energy, which, until now, has been established up to the single-particle electron $g$-factor \cite{kulik2021laughlin}. We found the zero-momentum spin-magnetograviton energy to be 1.6 meV, which is very close to the numerical value (1.57  meV) obtained in computer modeling of the finite electron system of 10 electrons. Furthermore, using the coherent anti-Stokes-Stokes scattering of light, we experimentally determined the density of the total saturation of the Laughlin liquid with spin-magnetogravitons, which makes it possible to study the new quasi-equilibrium state of electronic matter – the excited Laughlin liquid.

The authors are grateful to V. D. Kulakovskii, M. M. Glazov, and O. A. Grigorev for useful discussions. The work was supported by the Russian Science Foundation.

\end{document}